\documentclass[]{llncs}
\usepackage[T1]{fontenc}
\usepackage{graphicx}
\usepackage{amsmath} 
\usepackage{amsfonts}
\usepackage{booktabs}
\usepackage{multirow} 

\newcommand{\numUsers}{K}

\begin{document}

\title{Differentially Private Rankings via Outranking Methods and Performance Data Aggregation}
%
%
\author{Luis Del Vasto-Terrientes\inst{1}\orcidID{0000-0003-0483-8559}}
\authorrunning{Luis Del Vasto-Terrientes}
\titlerunning{Outranking-Based Differentially Private Rankings}

\institute{Universitat Rovira i Virgili, Departament d'Enginyeria Informàtica i Matemàtiques, Av. Paisos Catalans 26, Tarragona, 43007, Catalonia (Spain) 
\email{luismiguel.delvasto@fundacio.urv.cat}\\
}
\maketitle              
\begin{abstract}
Multiple-Criteria Decision Making (MCDM) is a sub-discipline of Operations Research that helps decision-makers in choosing, ranking, or sorting alternatives based on conflicting criteria. Over time, its application has been expanded into dynamic and data-driven domains, such as recommender systems. In these contexts, the availability and handling of personal and sensitive data can play a critical role in the decision-making process. Despite this increased reliance on sensitive data, the integration of privacy mechanisms with MCDM methods is underdeveloped.

This paper introduces an integrated approach that combines MCDM outranking methods with Differential Privacy (DP), safeguarding individual contributions' privacy in ranking problems. This approach relies on a pre-processing step to aggregate multiple user evaluations into a comprehensive performance matrix.

The evaluation results show a strong to very strong statistical correlation between the true rankings and their anonymized counterparts, ensuring robust privacy parameter guarantees.

\end{abstract}

\section{Introduction}
User preferences—encompassing opinions, likes, and choices—have been extensively studied across multiple disciplines. A prominent approach is Multiple-Criteria Decision Making (MCDM), a sub-discipline of Operations Research based on \textit{preference elicitation} and a \textit{constructive learning process}, where a decision-maker (DM) is central in the decision process. 
In contrast to data-driven approaches for automating preference modeling acquisition, such as Preference Learning (PL), MCDM allows for active participation in the definition of preferences based on performance rather than predicting them. 

Due to the rapid pace of technological advancement and the widespread use of distributed systems in modern IT environments, the development of hybrid model systems for decision-making has progressed significantly through the integration of MCDM methods and Machine Learning (ML) algorithms \cite{LIAO2023101970}. Multiple Criteria Recommender Systems \cite{Adomavicius2010} illustrate this integration by simultaneously optimizing multiple attributes to generate personalized recommendations. 

While MCDM methods have been successfully applied across many fields where sensitive data is crucial for decision-making, such as healthcare and tourism \cite{ALBAHRI2021101983,10.1007/s10844-015-0362-7,solano2024enhancing}, integrating privacy-preserving mechanisms into MCDM presents promising opportunities, particularly in data release scenarios where personal information may be at risk \cite{MADHUSUDHANAN2025104473}. In such scenarios, adversaries with side knowledge—such as known preferences—can infer specific sensitive contributions. This leakage can potentially compromise user privacy even in the absence of explicit identifiers. 

This study addresses the anonymization of user contributions in multi-criteria ranking problems for two well-known MCDM outranking methods: ELECTRE-III and PROMETHEE-II. The DM involved in the preference modeling is a stakeholder aiming to utilize available data to generate a ranking of alternatives for informed decision-making. It is essential to ensure that individual user contributions to this ranking remain unidentifiable. To achieve this, an integrated approach that incorporates Differential Privacy (DP)—the \textit{de facto} mechanism in the data privacy community due to its strong privacy guarantees—is evaluated. The results obtained reveal a significant statistical correlation between the true rankings and their anonymized counterparts, demonstrating the effectiveness of the proposed approach.

\subsection{Background \& Contributions}
The integration of traditional MCDM and DP for constructing anonymized preferences remains largely unexplored. However, some decision-making and DP related works highlight a shared focus. For example, \cite{torra2025differentially} introduces the differentially private Choquet integral, an advanced aggregation function known for its flexibility in handling interdependent and non-additive criteria, which has found wide application in MCDM. \cite{LAN2022115871} presents a differentially private approach for skyline queries that retrieves non-dominated objects from multidimensional datasets. 
Unlike MCDM, which typically involves evaluating a pre-defined set of alternatives, this work focuses on optimizing an unknown set of alternatives over multiple objectives, aligning with the principles of Multiple Objective Decision Making (MODM). 

DP-based ranking aggregation typically follows two approaches for combining multiple rankings into a collective one. The first involves users locally anonymizing their data—for instance, by injecting noise through Local DP (LDP) mechanisms \cite{yan2020private}—prior to sharing their information with potentially untrusted data curators. The second approach assumes the presence of a trusted data curator, who aggregates the original user preferences to produce noisy rankings \cite{shang2014}. However, these approaches do not consider the MCDM framework to rank alternatives.

This paper integrates the MCDM philosophy with DP for ranking problems by constructing a comprehensive performance matrix through the aggregation of user performances. DP is applied to this matrix to anonymize individual contributions before feeding the data into outranking methods. The DM plays a central role, ensuring the alignment of objectives and parameter setting. This approach is particularly valuable in scenarios where automated decision-making relies on dynamic user feedback and secure data sharing is essential. However, the strict mathematical definition of DP often reduces data utility in exchange for strong privacy guarantees. We explore Individual Differential Privacy (iDP), a relaxation of strict DP that enhances utility while maintaining robust data privacy.

The paper is organized as follows. Section \ref{sec:modelling} outlines the fundamental concepts of the decision-making process and introduces the ELECTRE-III and PROMETHEE-II outranking methods. Section \ref{sec:idp} introduces DP and iDP as privacy-preserving techniques. Section \ref{sec:idpmcdm} presents the approach for constructing anonymized rankings based on aggregated user feedback, and Section \ref{sec:results} details the results obtained from the experiments. Finally, Section \ref{sec:conclusions} presents the conclusions and suggests directions for future research.

\section{Problem modeling \& Outranking methods}
\label{sec:modelling}

The decision-making process models a problem by defining its structure, parameters, and functions using appropriate tools. This subjective model captures the DM's goals and comprises three essential elements:

\textbf{Alternatives:} $A = \{a, b, c, \dots\}$ represents the set of options available for selection. \textbf{Criteria:} $G = \{g_1, g_2, \dots, g_m\}$ evaluates alternatives, where $g_j(a)$ is the performance of $a \in A$ on $g_j \in G$, arranged in a performance matrix $M$ built for $A \times G$. \textbf{Preference system:} Defines relations such as preference ($aPb$), indifference ($aIb$)—describing a situation where two alternatives are equally acceptable—and incomparability ($aRb$), which corresponds to the absence of reasons to determine either $aPb$ or $aIb$. 

\subsection{ELECTRE-III and PROMETHEE-II}
\label{sec:outranking}

The outranking relation $S$, meaning ``is at least as good as'' is the core of outranking methods. For alternatives $(a,b)\in A\times A$, $a$ outranks $b$ if it performs well on enough important criteria and is not significantly outperformed by $b$ on any single criterion.

Outranking methods generally involve two primary steps: the construction of outranking relations and their subsequent exploitation. 

\subsubsection{ELECTRE-III}

\paragraph{Outranking relations} 

The outranking relation $aSb$ is based on two tests: the concordance test, where a majority of criteria should favor $aSb$, and the discordance test, where no criterion should strongly oppose $aSb$ if the concordance condition holds. It employs a non-compensatory model, such that poor performance on one criterion cannot be compensated by strong performance on another.

$G$ is considered as pseudo-criteria to handle imprecision or uncertainty relying on two thresholds: Indifference threshold $q_j(a)$: Below which alternatives are considered indifferent, and Preference threshold $p_j(a)$: Above which there is a clear preference for one alternative. Other required parameters are the relative importance of each criterion $w_j$ and the veto threshold $v_j(a)$, parameter used to express a strong opposition to an alternative being outranked by another.

The credibility $\rho(a, b)$ of outranking combines concordance and discordance indices, with a cutting level $\lambda$ commonly set between 0.5 and 1 \cite{FIGUEIRA2022986}. The concordance index $c(a, b)$ measures the coalition of criteria supporting $aSb$, while the discordance index $d_j(a, b)$ measures opposition strength. The overall credibility $\rho(a, b)$ is defined by:

\begin{equation} \label{eq:rankcred} \rho(a,b) =\begin{cases} c(a,b) & \text{ if } d_{j}(a,b) \leq c(a,b), \forall j \\ c(a,b) \prod\limits_{j \in J_{(a,b)}} \frac{1-d_{j}(a,b)}{1-c(a,b) } & \text{ otherwise } \end{cases} \end{equation}

where $J(a,b)$ is the set of criteria for which $d_{j}(a,b) > c(a,b)$.

\paragraph{Exploitation procedure}

The credibility matrix helps to create a partial preorder of alternatives via \textit{distillation}. This involves two steps: descending distillation, starting with the best alternatives, and ascending distillation, starting with the worst. The final partial preorder results from intersecting these descending and ascending complete preorders.

\subsubsection{PROMETHEE-II}

\paragraph{Outranking relations} 

Constructs a valued outranking relation using a preference index $P_j(a,b) \in [0, 1]$. 
The closer $P_j(a,b)$ is to 0, the greater the indifference between $a$ and $b$, while the closer the index is to 1, the stronger the preference of $a$ over $b$. Several functions have been proposed to model the degree of preferences, including linear preference with an indifference area and Gaussian criterion.

Assuming that for all pairs of alternatives $(a, b) \in A$, the preference indices $P_j(a, b)$ have been calculated, the overall preference $\Pi(a,b)$ is calculated taking into account a weight $w_j$ of each criterion $j$. This preference $\Pi(a, b)$ represents the weighted average of the partial preference functions $P_j(a,b)$. It is calculated as follows:

\begin{equation} \label{eq:prometheepi}
     \Pi (a, b) =  \frac{\sum^m_{j=1}w_j P_j(a, b)}{\sum^m_{j=1}w_j}\\
\end{equation}

\paragraph{Exploitation procedure} 

Alternatives are ranked using net flow, which balances two factors: leaving flow (how much an alternative outranks others) and entering flow (how much it is outranked). This net flow score determines the overall ranking, producing a final complete preorder.

\section{Differential Privacy}
\label{sec:idp}

The mathematical foundation of $\epsilon$-DP hinges on the principle that the presence or absence of any record in a dataset should not influence the protected outcome up to a factor exponential in $\epsilon$ \cite{dwork2006}. DP mitigates the risk of data disclosure by ensuring that analysis results remain nearly identical, even when comparing datasets that differ by just one record. DP is formalized as follows:

\begin{definition}
\label{def:diffpriv}
A DP mechanism $\kappa$ gives $\epsilon$-DP if, for all datasets \textit{$D_1$} and \textit{$D_2$} differing in at most one record (i.e., neighbor datasets) and all $S \subset$ Range($\kappa$), we have:
\begin{equation}
Pr(\kappa(D_1) \in S) \leq exp(\epsilon) Pr(\kappa(D_2) \in S)
\end{equation}
\end{definition}

$\epsilon$ controls the noise addition to the value, where $\epsilon = 0$ ensures the highest possible privacy guarantees, while $\epsilon = \infty$ provides no privacy guarantees. 

The Laplace mechanism is the most common method for achieving DP, adding noise drawn from the Laplace distribution. The magnitude of the noise is determined by the sensitivity of the original data to changes, reflecting the potential distortion.

DP relies on the global sensitivity, the maximum variability between neighboring datasets in the same domain  $\mathcal{D}$, assuming the most adversarial condition:

\begin{definition}[Global sensitivity]
\label{def:global_sensitivity}
Let $f$ be a function that is evaluated at datasets in $\mathcal{D}$
and returns values in $\mathbb{R}^{u}$. The global sensitivity of $f$ 
over $\mathcal{D}$ is defined as:
\begin{equation}
\Delta f = \max_{\begin{array}{c}
    {\scriptstyle D_1,D_2\in\mathcal{D}}\\
    {\scriptstyle d(D_1,D_2)=1}
\end{array}} \left\Vert f(D_1)-f(D_2) \right\Vert_1,
\end{equation}
where $d(D_1,D_2)$ means that datasets $D_1$ and $D_2$ differ in one record. 
\end{definition}

One of the most interesting properties of DP is composability, which is considered for a sequence of DP mechanisms, such as consecutive queries executed over the same database.

\begin{definition}
\label{def:sequential}
\emph{Sequential composition}. For $\epsilon$-differentially private mechanisms $\kappa_1$ and $\kappa_2$ accessing \emph{non-independent} datasets $D_1$ and $D_2$ respectively, the combination of their outputs provides ($\kappa_1$ + $\kappa_2$)-DP.
\end{definition}

\begin{definition}
\label{def:parallel}
\emph{Parallel composition}. For $\epsilon$-differentially private mechanisms $\kappa_1$ and $\kappa_2$ accessing \emph{disjoint and independent} datasets $D_1$ and $D_2$ respectively, the combination of their output provides $\max(\kappa_1, \kappa_2)$-DP.
\end{definition}

\subsection{Individual Differential Privacy}
\label{sec:dpirm}
iDP \cite{soria2017individual} calibrates noise addition based on the \textit{actual} dataset (rather than \textit{any} possible dataset), reducing information loss compared to strict DP, which follows a worst-case scenario approach. It satisfies the following condition for the actual dataset $D_1$ and any neighbor dataset $D_2$ differing by one record:
\begin{equation}
exp(-\epsilon) Pr(\kappa(D_2) \in S) \leq Pr(\kappa(D_1) \in S) \leq exp(\epsilon) Pr(\kappa(D_2) \in S)
\end{equation}

To achieve iDP, noise is calibrated to local sensitivity, defined for a function $f$ on dataset $D_1$ as:

\begin{definition}[Local sensitivity] \label{def:localsensitivity} Let \( D_1 \) be the actual dataset and let \(\mathcal{D}\) be a collection of datasets, where \( D_2 \) is any neighboring dataset. The local sensitivity of a function $f$ in $D_1$ is \begin{equation} LS_{f}(D_1)= \max_{D_2 : d(D_2, D_1) = 1} \| f(D_2) - f(D_1) \|_1, \end{equation} where $d(D_2, D_1)$ means that dataset $D_2$ differs from $D_1$ in one record. \end{definition}

\section{Outranking Methods and DP Integration}
\label{sec:idpmcdm}

This section presents a privacy-preserving approach for multi-criteria ranking applying outranking methods. It addresses scenarios in which multiple users collaboratively construct a comprehensive performance matrix for a finite set of alternatives. The approach involves three key steps, which are presented below.

\subsection{Aggregation of User Performances: 
Constructing $OM$}
\label{sec:opm}
Let $M$ denote a performance matrix that contains personalized single-user evaluations over many alternatives. Suppose a system has access to multiple such performance matrices and needs to rank the top 5 alternatives. To accomplish this, the system constructs $OM$ from the performance scores provided by users. $OM$ then serves as an overall performance matrix that provides a comprehensive assessment of alternatives by integrating diverse perspectives and mitigating individual biases. Evaluations may include any supported metric, whether quantitative or qualitative, that measures the performance of an alternative. For simplicity, we use the same notation for $OM$ as that used for $M$ in Section \ref{sec:modelling}.

Given that $OM$ serves as a general benchmark derived from individual evaluations, we compute $g_j(a)$ using an aggregation operator. In this work, the simple arithmetic mean is used.

\begin{subsubsection}{Simple Arithmetic Mean}
\label{sec:sma}
The simple arithmetic mean (SAM) is one of the most basic models of group judgment and considers each member as equally influential. We compute the arithmetic mean of individuals' performance values as follows:

\begin{equation}
SAM_{g_j(a)} = \frac{1}{\numUsers} \sum_{k=1}^{\numUsers} g_j^{k}(a)
\end{equation}

Let $\numUsers(a)$ be the set of users evaluating alternative $a$, and let $\numUsers = |\numUsers(a)|$. The average performance of alternative $a$ on criterion $j$, computed from these $\numUsers$ evaluations, is denoted by $SAM_{g_j(a)}$.

In this work, we assume that for alternatives $a$ and $b$, $|\numUsers(a)| = |\numUsers(b)| = \numUsers$ and $\numUsers(a) \cap \numUsers(b) = \emptyset$. Thus, $OM$ is constructed from a disjoint set of $\numUsers \times |A|$ users, meaning that each user contributes an evaluation for a single alternative within $M$.

The resulting $OM$ undergoes noise addition via a DP randomized mechanism. This strategy enhances utility after applying DP noise, as explained in more detail in the next section.
\end{subsubsection}

\subsection{Differentially Private $OM$}
\label{sec:dpopm}

According to Section \ref{sec:opm}, alternatives in $OM$ are constructed from disjoint sets of users. Therefore, each user contributes performance data for only one alternative per ranking request.

This approach improves the utility by taking advantage of the parallel composition property, which allows the full privacy budget $ \epsilon $ to be independently allocated to each alternative (see Definition \ref{def:parallel}). Within each alternative, sequential composition is then used to distribute $ \epsilon $ among $G$. This distribution can be uniform or tailored to the DM's preferences, for example, by assigning more $ \epsilon $ to a more sensitive criterion. When distributed evenly, each performance value $ g_j(a) $ is allocated a privacy budget of $ \frac{\epsilon}{|G|} $. Following this strategy, an $ \epsilon $-DP $ OM^* $ is constructed.

User contributions are then anonymized by injecting calibrated noise into each aggregated performance $ g_j(a) $ in $OM$, producing its masked version $ g_j^*(a) $ in $OM^*$. Consequently, $ a^* $ (masked version of $a$) is computed by independently and consecutively querying all aggregated performance values $ g_j(a)$ for $\forall j \in \{1, 2, ..., m \} $.

In DP, the sensitivity quantifies the maximum change in the output of a function when a single data point is modified. Therefore, the scale parameter for the applied noise distribution (e.g., Laplace or Gaussian) is determined by the aggregation operator. 

The global sensitivity $\Delta SAM_{g_j(a)}$ is

\begin{equation}
\Delta SAM_{g_j(a)} = \frac{\max D^{g_j} - \min D^{g_j}}{\numUsers}
\end{equation}

where $\max D^{g_j}$ and $\min D^{g_j}$ represent the maximum and minimum possible values in the criterion domain $D^{g_j}$, respectively.

On the other hand, the local sensitivity $LS_{SAM{g_j(a)}}$ is 

\begin{equation}
    \label{prop:sens_first}
    LS_{SAM{g_j(a)}} = \frac{\max\{\max D^{g_j}-\min(g_j(a)),\,\max(g_j(a))-\min D^{g_j}\}}{\numUsers}
\end{equation}

where $\max\left( g_j(a) \right)$ and $\min\left( g_j(a) \right)$ are functions that return, respectively, the maximum and minimum performance values in alternative $a$ and criterion $j$ across $\numUsers$; and the maximum and minimum possible values in the criterion domain $D^{g_j}$ are denoted by $\max D^{g_j}$ and $\min D^{g_j}$, respectively.

The aggregation step using $SMA$ effectively reduces the sensitivity to noise by diluting its impact on the final output through $\numUsers$. In addition to the privacy protection provided by DP, the system provides uncertainty about whether a particular individual contributed or not to $OM$, further improving privacy.

\subsection{Applying Outranking Methods to $OM^*$}

$OM^*$—which masks the contribution of individual users in the resulting preorder—is computed to serve as input for the outranking methods. Therefore, the outranking methods are implemented as outlined in Section \ref{sec:modelling}.

\section{Empirical Results}
\label{sec:results}

This section presents the results of the empirical evaluation of the proposed approach for integrating DP and iDP with the ELECTRE III and PROMETHEE II outranking methods. The evaluation was conducted using two datasets.

\begin{itemize}
    \item A synthetic data generator was developed to simulate user scores ranging between [0-100], a scale commonly used by many platforms such as Metacritic to reflect critical reception of media. The synthetic data generator assigns scores to criteria based on predefined alternative categories: \textbf{High-scoring:} First 33\% of alternatives tend to score 70–85 per criterion. \textbf{Mid-scoring:} Next 33\% tend to score 50–75.
\textbf{Low-scoring:} Remaining alternatives tend to score 30–55. 

The values within each score range concentrate 70\% of the probability distribution. Consequently, the score distribution within each range does not restrict the selection of ratings outside the range.

    \item BeerAdvocate is a dataset that contains 1.5 million user ratings of beers according to 4 criteria: aroma, appearance, palate, and taste. Ratings range from 1 to 5 in 0.5-point increments, giving users nine possible scoring options.
\end{itemize}

The assessment measures the information loss of $OM^*$ relative to $OM$, followed by an analysis of the correlation between the true and anonymized rankings, produced by applying the outranking methods to $OM$ and $OM^*$, respectively. Several predefined $(\epsilon, K)$ pairs are tested, and the reported results are averaged over 50 iterations to account for the stochastic nature of DP.

Information loss is quantified using the mean absolute error (MAE), a widely used metric for numerical data. MAE is calculated as the sum of absolute differences between the corresponding elements of two matrices, divided by the number of elements compared. Utility is maximized when MAE equals zero, indicating identical matrices—and thus, no privacy protection.

The Spearman rank correlation coefficient $r_s$ is calculated to assess the correlation between the true and anonymized rankings. Analyzing DP and iDP results under the same $(\epsilon, K)$ offers valuable insight into the influence of global and local sensitivity on ranking utility. Since iDP noise calibration depends on the actual aggregated data values, it is expected to preserve utility more effectively than DP. The masking process applies Laplace noise.

The tests were conducted on a Windows 11 Home PC with an Intel i7-1355U CPU @5.00 GHz and 16 GB DDR5 RAM.
The OpenDP library, a comprehensive toolkit designed for implementing DP
, and the pyDecision library
, which offers a wide set of MCDM methods, have both been used for this evaluation.

\subsubsection{Synthetic dataset}

A ranking scenario involving 12 available alternatives evaluated on 6 criteria is considered. The main parameter values are $\epsilon$=[0.1, 0.5, 1] and $K$=[10, 25, 50, 100, 250, 400]. For the 6 criteria, differences within 3 units ($q_j(a) = 3$) are negligible, while a 25-unit difference ($v_j(a) = 25$) acts as veto threshold in ELECTRE III. Preference thresholds vary between 10 and 20 units depending on the criterion ($p_j(a)$=[10, 20, 20, 10, 20, 10]). Weights ($W$) are set as [0.2, 0.15, 0.1, 0.25, 0.1, 0.2], and PROMETHEE II applies the level criterion as preference function.

\paragraph{Synthetic data results}
\label{sec:electretests}

A comprehensive comparison of the information loss between DP and iDP is presented in Table \ref{table:synthetic}.

\begin{table}[ht]
\centering
\caption{MAE results for the synthetic dataset}
\label{table:synthetic}
\begin{tabular}{|c|cc|cc|cc|cc|cc|cc|}
\hline
\multirow{2}{*}{\textbf{$\epsilon / K$}} & \multicolumn{2}{c|}{\textbf{10}} & \multicolumn{2}{c|}{\textbf{25}} & \multicolumn{2}{c|}{\textbf{50}} & \multicolumn{2}{c|}{\textbf{100}} & \multicolumn{2}{c|}{\textbf{250}} & \multicolumn{2}{c|}{\textbf{400}} \\
\cline{2-13}
 & \textbf{DP} & \textbf{iDP} & \textbf{DP} & \textbf{iDP} & \textbf{DP} & \textbf{iDP} & \textbf{DP} & \textbf{iDP} & \textbf{DP} & \textbf{iDP} & \textbf{DP} & \textbf{iDP} \\
\hline
\textbf{0.1} & 48.04 & 47.17 & 45.16 & 43.81 & 40.63 & 38.93 & 34.21 & 30.09 & 20.70 & 18.63 & 14.24 & 12.84 \\
\hline
\textbf{0.5} & 41.01 & 37.45 & 30.43 & 27.11 & 21.22 & 17.43 & 11.78 & 9.98  & 4.84  & 3.91  & 2.99  & 2.74 \\
\hline
\textbf{1}   & 33.18 & 29.03 & 20.83 & 16.96 & 11.67 & 9.88  & 6.18  & 5.03  & 2.31  & 2.09  & 1.48  & 1.28 \\
\hline
\end{tabular}
\end{table}

The results show that iDP consistently achieves lower MAE values than DP across all fixed $(\epsilon,K)$ settings, particularly for lower $K$. As more data is aggregated, local sensitivity tends to approach global sensitivity, since the likelihood of encountering minimum and maximum domain values increases in the generated dataset. Furthermore, increasing both $\epsilon$ and $K$ steadily reduces the distortion in the anonymized output. This highlights the importance of carefully tuning the parameters to achieve an optimal balance between privacy and utility.

\begin{figure}
    \centering
    \includegraphics[width=1\linewidth]{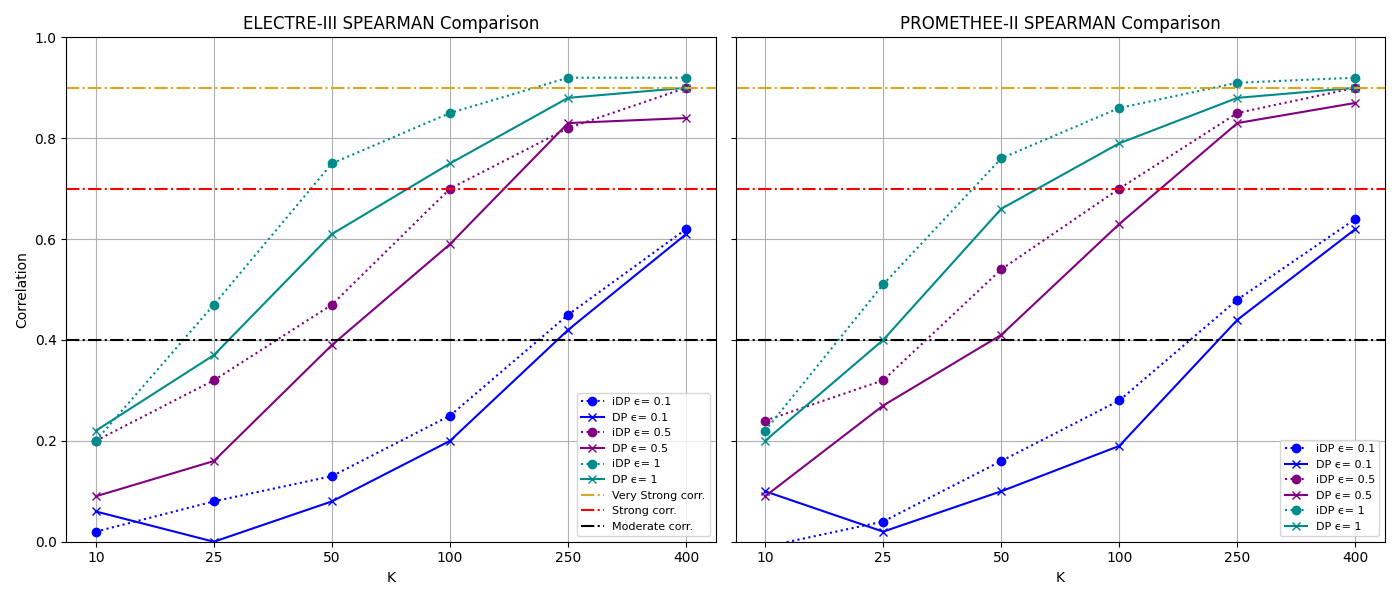}
    \caption{Synthetic data results}
    \label{fig:syntheticresults}
\end{figure}

Fig. \ref{fig:syntheticresults} presents the averaged $r_s$ obtained for the predefined $(\epsilon, K)$ values. Notice that the thresholds indicating moderate, strong, and very strong correlations are plotted on the graph to better understand the level of association between the true and anonymized rankings. We follow the conventional guideline provided in \cite{schober2018correlation}. Correlation strength is negligible to weak for $r_s$ in [0.00, 0.4), moderate in [0.4, 0.7), strong in [0.7, 0.9), and very strong in [0.9, 1], reflecting near-perfect relationships.

In line with the observations in Table \ref{table:synthetic}, for a fixed $K$, increasing $\epsilon$ leads to better $r_s$ for both DP and iDP due to the reduced amount of added noise (weaker privacy guarantees). In addition, the strength of $r_s$ increases for fixed $\epsilon$ and growing $K$ due to lower noise sensitivity when more data is aggregated to $g_j(a)$. Consequently, lower $\epsilon$ values (stronger privacy guarantees) can be applied without compromising utility as $K$ grows.

For fixed $(\epsilon, K)$, better $r_s$ values are typically achieved for iDP compared to DP, which often leads to over-perturbation and, consequently, to more impact on the outputs. For example, based on the results obtained, strong $r_s$ is achieved for iDP for $(\epsilon=1, K=50)$ for both outranking methods, while for strict DP, a strong correlation is achieved for $(\epsilon=1, K=100)$. 

\subsubsection{BeerAdvocate dataset}
A ranking scenario involving 20 alternatives evaluated on 4 criteria—aroma, appearance, palate, and taste—is considered. 
The main parameter values used are $\epsilon$ = [0.1, 0.5, 1] and $K$=[25, 50, 75, 100, 200]. The specific parameter values for the outranking methods are set as follows: 
differences within 0.5 units ($q_j(a) = 0.5$) are considered negligible, while a 4-unit difference serves as veto ($v_j(a) = 4$). The preference threshold is set at 1 unit ($p_j(a) = 1$). In addition, all criteria are assumed to be equally important, and the level criterion is chosen as the preference function for PROMETHEE-II. 

In each iteration, 20 alternatives are randomly selected from those with at least $K$ evaluations. For alternatives with more than $K$ evaluations, selection prioritizes those closest to the average review profile, which is determined by computing the mean scores in all criteria in all evaluations. The most representative alternatives are then identified by minimizing their Euclidean distance from this profile.

The results of the information loss are presented in Table \ref{tab:mae_beercomparison}.

\begin{table}[ht]
\centering
\caption{MAE results for the BeerAdvocate dataset}
\label{tab:mae_beercomparison}
\begin{tabular}{|c|cc|cc|cc|cc|cc|cc|}
\hline
\multirow{2}{*}{\textbf{$\epsilon / K$}} & \multicolumn{2}{c|}{\textbf{10}} & \multicolumn{2}{c|}{\textbf{25}} & \multicolumn{2}{c|}{\textbf{50}} & \multicolumn{2}{c|}{\textbf{75}} & \multicolumn{2}{c|}{\textbf{100}} & \multicolumn{2}{c|}{\textbf{200}} \\
\cline{2-13}
 & \textbf{DP} & \textbf{iDP} & \textbf{DP} & \textbf{iDP} & \textbf{DP} & \textbf{iDP} & \textbf{DP} & \textbf{iDP} & \textbf{DP} & \textbf{iDP} & \textbf{DP} & \textbf{iDP} \\
\hline
\textbf{0.1} & 1.84 & 1.83 & 1.65 & 1.63 & 1.43 & 1.31 & 1.23 & 1.13 & 1.08 & 0.98 & 0.68 & 0.60 \\
\hline
\textbf{0.5} & 1.41 & 1.35 & 0.95 & 0.82 & 0.59 & 0.49 & 0.41 & 0.36 & 0.31 & 0.26 & 0.17 & 0.14 \\
\hline
\textbf{1}   & 1.01 & 0.94 & 0.59 & 0.49 & 0.32 & 0.26 & 0.21 & 0.18 & 0.16 & 0.13 & 0.08 & 0.07 \\
\hline
\end{tabular}
\end{table}

Following the same pattern of decreasing information loss shown in Table \ref{table:synthetic}, utility is better preserved as the values of $\epsilon$ and $K$ increase. In addition, iDP provides less information loss in comparison to strict DP for the same $(\epsilon, K)$ values.

\begin{figure}[ht]
    \centering
    \includegraphics[width=1\linewidth]{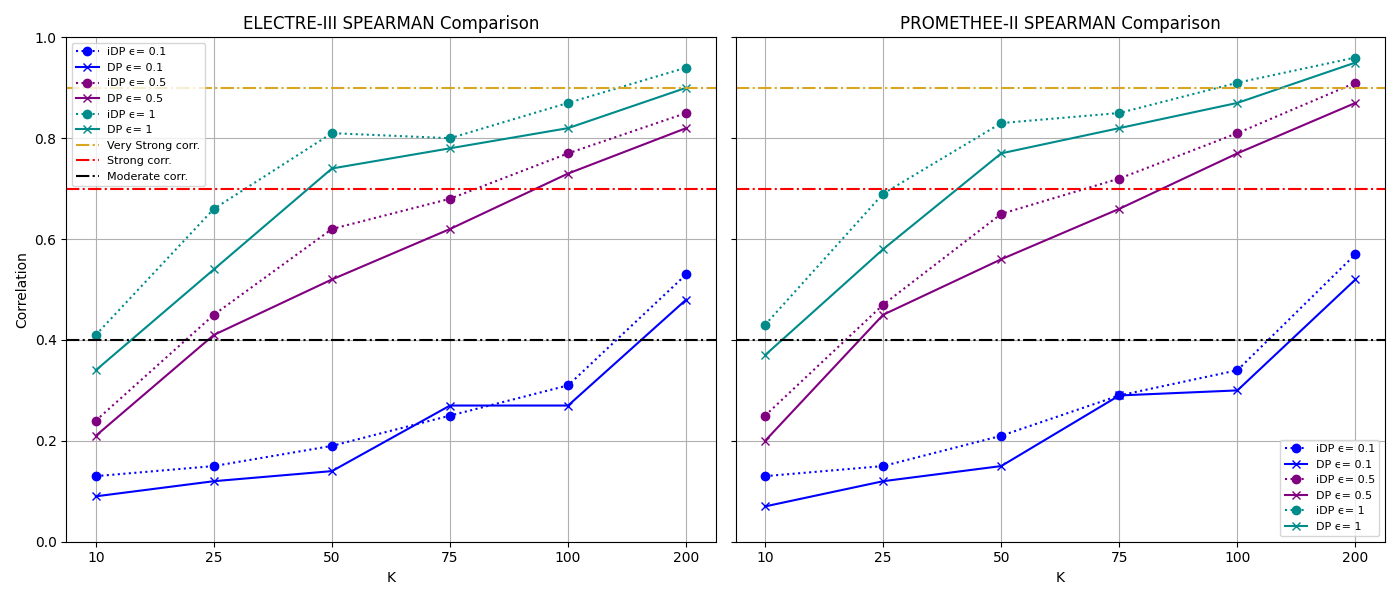}
    \caption{BeerAdvocate Results}
    \label{fig:beeradvresults}
\end{figure}

Fig.~\ref{fig:beeradvresults} shows the consistent trend of $r_s$ with those observed in the synthetic dataset. For a fixed $K$, increasing $\epsilon$ results in better correlation values. Likewise, for a fixed $\epsilon$, increasing $K$ enhances ranking utility due to the reduction of noise sensitivity. Finally, for a fixed setting $(\epsilon, K)$, the iDP mechanism typically slightly outperforms strict DP, confirming that leveraging local sensitivity enables better utility preservation in privacy-preserving ranking.

\label{sec:beerresults}

\section{Conclusions and Future Work}
\label{sec:conclusions}

The exponential growth of data collection and processing has accelerated the development of hybrid decision-making systems that combine multiple techniques from different fields, such as ML and MCDM, to leverage their respective strengths. However, this vast amount of personal data also underscores the need to prioritize individual privacy protection.

In this work, we explore the integration of DP as a privacy mechanism with two MCDM outranking methods, ELECTRE-III and PROMETHEE-II. The approach relies on a pre-processing step to construct an overall performance matrix, where individual user contributions are masked. The conducted experiments indicate strong correlations for safe $\epsilon$ values using DP and iDP. iDP requires less data availability to achieve higher utility, which makes iDP advantageous in scenarios where systems have limited data for aggregation. Future research may explore alternative approaches to further enhance utility in such cases, ensuring that moderate to strong correlations are consistently preserved while maintaining robust privacy guarantees.

\begin{credits}
\subsubsection{\ackname} Partial support for this work has been received from MCIN/AEI/ 10.13039/501100011033 and ``ERDF A way of making Europe'' under grant PID2021-123637NB-I00 ``CURLING'', and by Project HERMES funded by the European Union NextGenerationEU/PRTR via INCIBE.

\end{credits}

\bibliographystyle{splncs04}
\bibliography{source}
\end{document}